\documentclass{article}

\usepackage{microtype}
\usepackage{graphicx}
\usepackage{subfigure}
\usepackage{array}
\usepackage{booktabs} % for professional tables
\graphicspath{{figures/}}
\usepackage{hyperref}

\usepackage[accepted]{sysml2019}

\sysmltitlerunning{Accelerator-aware Neural Network Design using AutoML}

\begin{document}

\twocolumn[
\sysmltitle{Accelerator-aware Neural Network Design using AutoML}

% It is OKAY to include author information, even for blind
% submissions: the style file will automatically remove it for you
% unless you've provided the [accepted] option to the sysml2019
% package.

% List of affiliations: The first argument should be a (short)
% identifier you will use later to specify author affiliations
% Academic affiliations should list Department, University, City, Region, Country
% Industry affiliations should list Company, City, Region, Country

% You can specify symbols, otherwise they are numbered in order.
% Ideally, you should not use this facility. Affiliations will be numbered
% in order of appearance and this is the preferred way.
\sysmlsetsymbol{equal}{*}

\begin{sysmlauthorlist}
\sysmlauthor{Suyog Gupta}{goog}
\sysmlauthor{Berkin Akin}{goog}
\end{sysmlauthorlist}

\sysmlaffiliation{goog}{Google Inc., Mountain View, CA}

\sysmlcorrespondingauthor{Suyog Gupta}{suyoggupta@google.com}
\sysmlcorrespondingauthor{Berkin Akin}{bakin@google.com}

% You may provide any keywords that you
% find helpful for describing your paper; these are used to populate
% the "keywords" metadata in the PDF but will not be shown in the document
\sysmlkeywords{Machine Learning, SysML}

\vskip 0.3in
\begin{abstract}
While neural network hardware accelerators provide a substantial amount of raw compute throughput, the models deployed on them must be co-designed for the underlying hardware architecture to obtain the optimal system performance. We present a class of computer vision models designed using hardware-aware neural architecture search and customized to run on the Edge TPU, Google's neural network hardware accelerator for low-power, edge devices. For the Edge TPU in Coral devices, these models enable real-time image classification performance while achieving accuracy typically seen only with larger, compute-heavy models running in data centers. On Pixel 4's Edge TPU, these models improve the accuracy-latency tradeoff over existing SoTA mobile models.
%Neural network hardware accelerators and the models deployed on them must be co-designed to obtained optimal system performance. In this work, we present a class of neural network models for computer vision tasks customized to run on the Edge TPU, Google's neural network hardware accelerator for edge devices. 
\end{abstract}
]

% this must go after the closing bracket ] following \twocolumn[ ...

% This command actually creates the footnote in the first column
% listing the affiliations and the copyright notice.
% The command takes one argument, which is text to display at the start of the footnote.
% The \sysmlEqualContribution command is standard text for equal contribution.
% Remove it (just {}) if you do not need this facility.

\printAffiliationsAndNotice{}  % leave blank if no need to mention equal contribution
%\printAffiliationsAndNotice{\sysmlEqualContribution} % otherwise use the standard text.

\section{Introduction}
On-device machine learning (ML) strives to bring privacy-preserving, always-available, and responsive intelligence to compute platforms that may be limited in terms of compute and power resources. Enabling on-device ML on resource-constrained devices has spurred the development of algorithmically-efficient neural network architectures as well as a myriad of specialized hardware accelerators architected to efficiently execute the kernels commonly found in deep neural networks. 
These hardware accelerators exhibit quite a bit of diversity in terms of their programming models, compute capabilities, memory organization and it is unlikely that the same neural network architecture can map efficiently across these different hardware platforms. Put differently, neural network architectures must be aware of the target hardware architecture in order to optimize the overall system performance and energy efficiency.

Meanwhile there is an increasing trend in the deep learning community to employ automated neural architecture search (NAS) methods to design models, 
representing a shift from the conventional approach of hand tuning the model architectures. 
Much of the early work in NAS relied on the number of multiply-accumulate (MAC) operations performed during inference or the number of trainable parameters as a proxy for model's latency \cite{nasnet}. 
Realizing that the MAC count does not always correlate well with the measured latency on a real device, recent work such as MNASNet\cite{mnasnet}, ProxylessNAS\cite{proxylessnas}, 
and FBNet \cite{fbnet} first build a model to estimate the network's latency on a target hardware. 
This latency model is then used to navigate the space of candidate architectures. 
We extend these NAS frameworks to search for computer vision models customized for the different instantiations of Google's Edge TPU neural network hardware accelerator architecture: Edge TPU in the USB/PCI-e attached Coral devices\footnote{\url{https://coral.ai/products/}} and in the Pixel 4 smartphone\footnote{\url{https://store.google.com/product/pixel_4}}. 
We pay special attention to the design of the search space used for sampling the candidate neural network architectures. 
In particular, we augment the search space with building blocks known to achieve high overall utilization on the Edge TPU architecture. 
In addition, we prohibit the use of operations incompatible with the production software stack, 
thereby yielding models that are readily deployed on the target devices. 

This accelerator-aware NAS is used to discover efficient image classification models for the Edge TPU: 
(i) EfficientNet-EdgeTPU runs nearly 10x faster compared with Resnet50 \cite{resnet} on the Edge TPU in Coral devices while achieving higher classification accuracy.
(ii) For the Edge TPU in Pixel 4 the search for smaller, low-latency models produced MobilenetEdgeTPU which achieve 75.6\% top-1 accuracy while reducing latency by 30\% over MobilenetV3.
We also released the training code and pretrained versions for these models\footnote{\url{https://github.com/tensorflow/models/tree/master/research/slim/nets/mobilenet}},\footnote{\url{https://github.com/tensorflow/tpu/tree/master/models/official/efficientnet/edgetpu}}.
%This accelerator-aware NAS is used to discover efficient image classification models for the Edge TPU in Coral devices. 
%The resulting model, called Efficient-EdgeTPU, runs nearly 10x faster compared with Resnet50 \cite{resnet} on Edge TPU while achieving higher classification accuracy. For the Edge TPU in Pixel4 the search for smaller, low-latency models produced MobilenetEdgeTPU which achieve 75.6\% top-1 accuracy while reducing latency by ~30\% over MobilenetV3. 
\section{Methodology}

\begin{figure}[t]
    \centering
    \includegraphics[width=\linewidth]{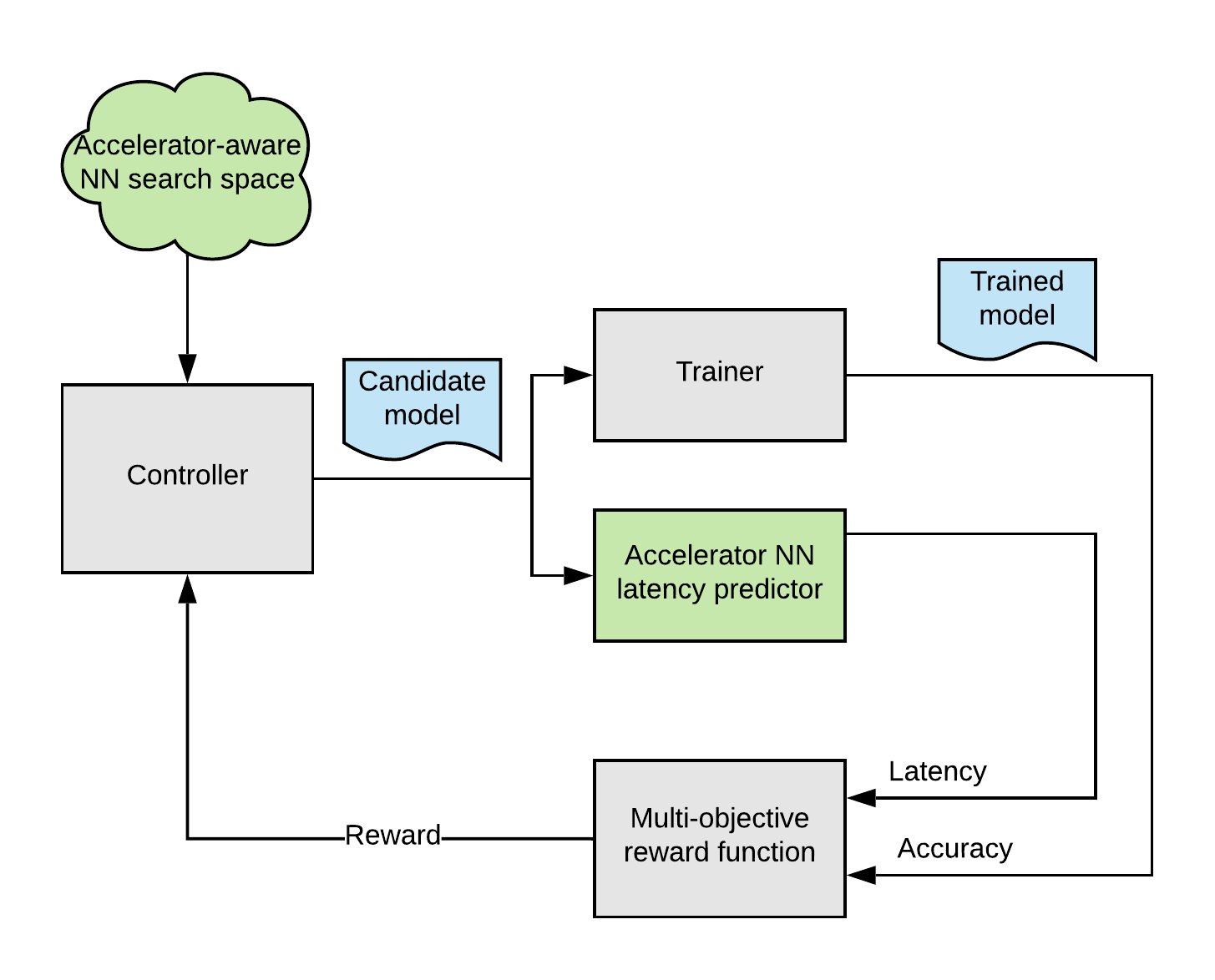}
    \vskip -0.1in
    \caption{Accelerator performance predictor added into the loop with neural architecture search for discovering accelerator optimized models.}
    \label{fig:NAS}
    \vskip -0.1in
\end{figure}

A typical neural architecture search framework consists of the following modules: a controller that samples from search space of all possible architectures, a trainer that trains the models on some dataset to arrive at an accuracy metric, an objective function that scores the candidate model to help the controller navigate the search space. As shown in Figure~\ref{fig:NAS}, this framework can be extended to search for accelerator-optimized models by integrating a ``latency model" that returns a model's latency when running on the target accelerator, and an objective function that jointly optimizes the model latency and accuracy. 

\subsection{Accelerator Performance Modeling}
Model latency and accuracy are the two components of the multi-objective reward function in hardware-aware NAS. Hence, projecting the latency of a candidate model running on the underlying hardware both quickly and accurately is a key challenge. We have leveraged a variety of performance evaluation strategies with inherent trade-offs between estimation speed and accuracy.

As the utilization of the hardware resources highly depend on the computation characteristics of the operations (e.g., memory accesses, data reuse, compute intensity, etc.) employed in the neural network models, simply using the number of MACs or parameters in the models as a proxy for the latency can be very misleading. For example, as shown in \cite{mnasnet}, MobileNet \cite{howard2017mobilenets} and NASNet \cite{nasnet} have similar FLOPS (575M vs. 564M), but the model latencies can be significantly different (113ms vs. 183ms on Pixel 1). On the other hand, building a test harness of real devices to measure the latencies of several models explored in NAS poses several scalability challenges.

\begin{figure}[t]
    \centering
    \includegraphics[width=\linewidth]{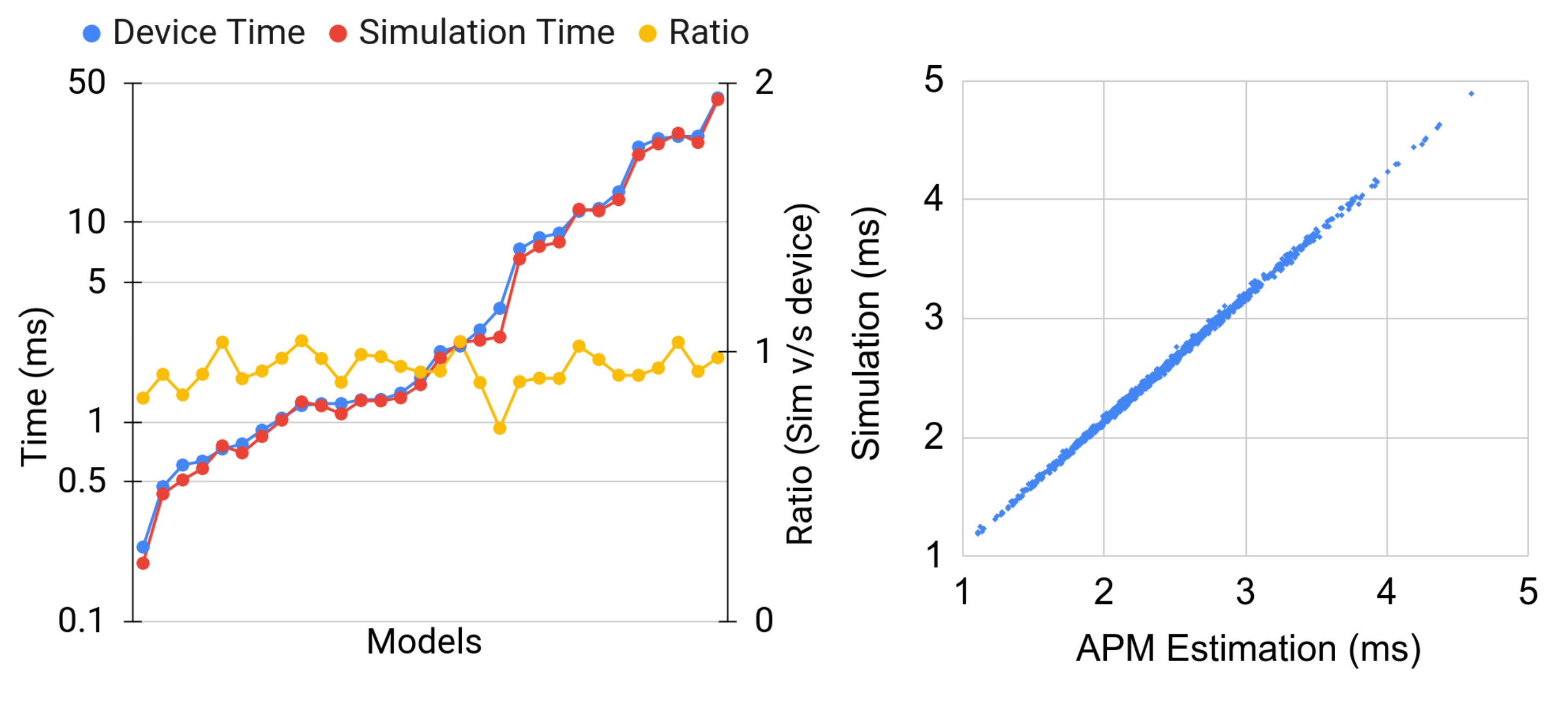}
    \vskip -0.1in
    \caption{Left: Simulated vs. real device latencies for a mix of ML models. Right: Simulated vs. APM estimated latencies for 1000 randomly sampled models from the search space (RMSE=160us)}
    \label{fig:sim-vs-device}
    \vskip -0.15in
\end{figure}

To address the challenges of real-device measurements, we used a cycle-accurate Edge TPU performance simulator to estimate the latencies of the candidate models. Our simulator faithfully models most of the key subsystems to evaluate full models under a few minutes while providing a very close proxy for the real device (see Figure~\ref{fig:sim-vs-device}).

Cycle-level simulation provides accurate latency estimations and a few minutes of turnaround time for the estimation can be hidden behind a costly training of a proxy task for some NAS approaches \cite{mnasnet}. However, some NAS algorithms based on weight sharing in a super-network \cite{tunas,proxylessnas} may require a much faster latency estimation. For that purpose, we have developed an analytical performance model (APM) for Edge TPU that provides latency estimations in the order of milliseconds. APM uses a roofline model where the roofline constructed from the peak memory bandwidth and compute throughput is enhanced with multiple ceilings that capture some of the key performance limiters such as internal data buses, local memory units, etc. \cite{roofline}. Based on 1000 randomly sampled models from our image classification search space, APM provides reasonably accurate estimations (160us RMSE) with 3-4 orders of magnitude speedup in estimation time compared to cycle-level simulation (see Figure~\ref{fig:sim-vs-device}).

Estimating model latency, regardless of using analytical model or cycle-level simulation, is coupled with several components of the Edge TPU software stack such as model transformation and compilation. We expose our evaluation toolchain as a remote service and deploy it as a latency estimation server. The service interface facilitates the integration of Edge TPU software stack and the simulator/APM into the NAS framework by using remote procedure calls (RPC). Moreover, it provides an efficient way to scale-up the latency estimation where several candidate models are explored in parallel.

\subsection{Search Space Engineering}

Domain-specific accelerators owe much of their performance and energy efficiency gains over general purpose processors to the customization of their compute units and datapaths for specific computation patterns. This also implies that not all operations will achieve similar computational efficiencies when mapped to such accelerators. As a result, crafting the search space to include building blocks that are known to improve hardware utilization as well as excluding incompatible operations becomes a critical component in arriving at accelerator-optimized models.

\begin{figure}[t]
    \centering
    \includegraphics[width=\linewidth-2em]{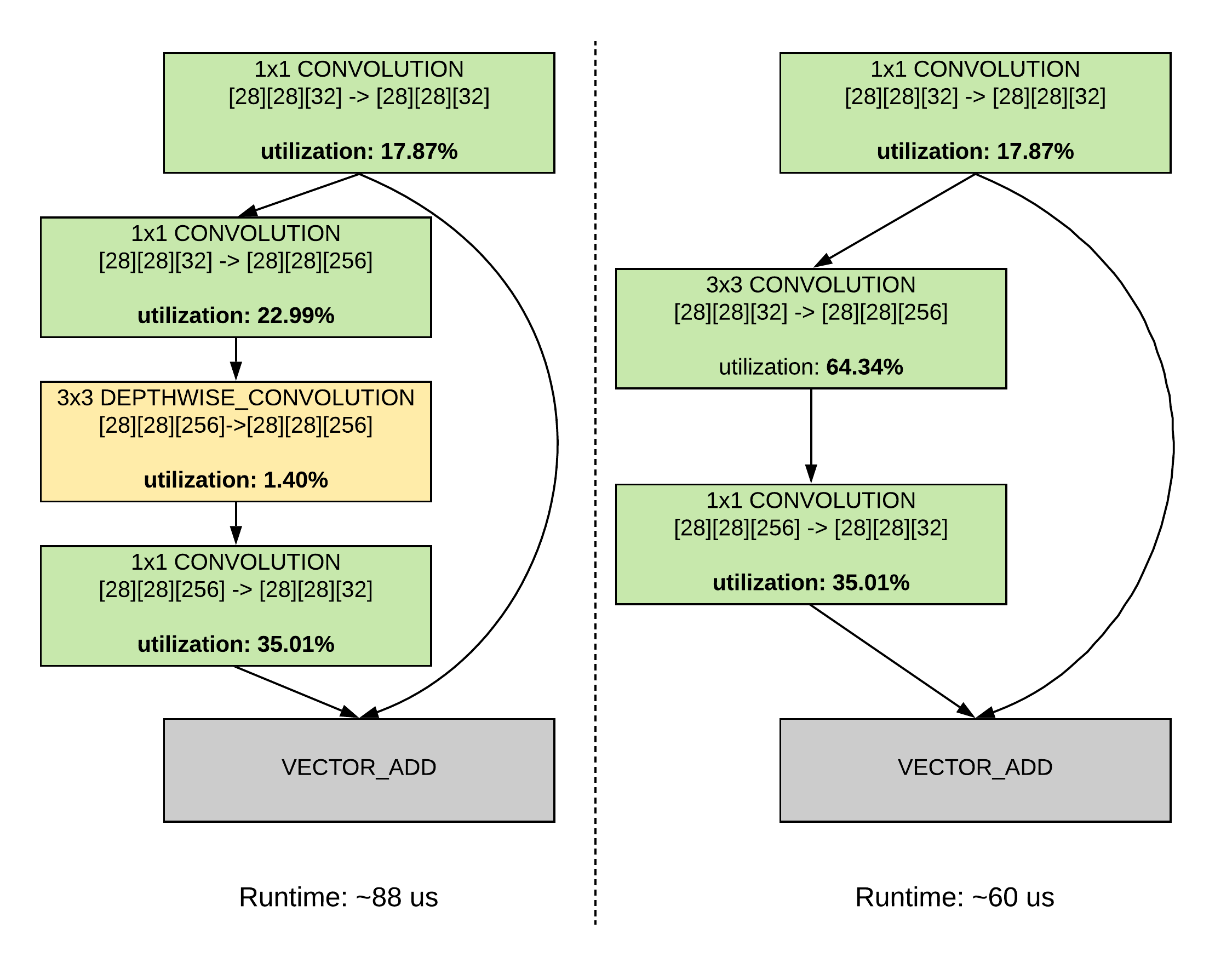}
    \vskip -0.2in
    \includegraphics[width=\linewidth-2em]{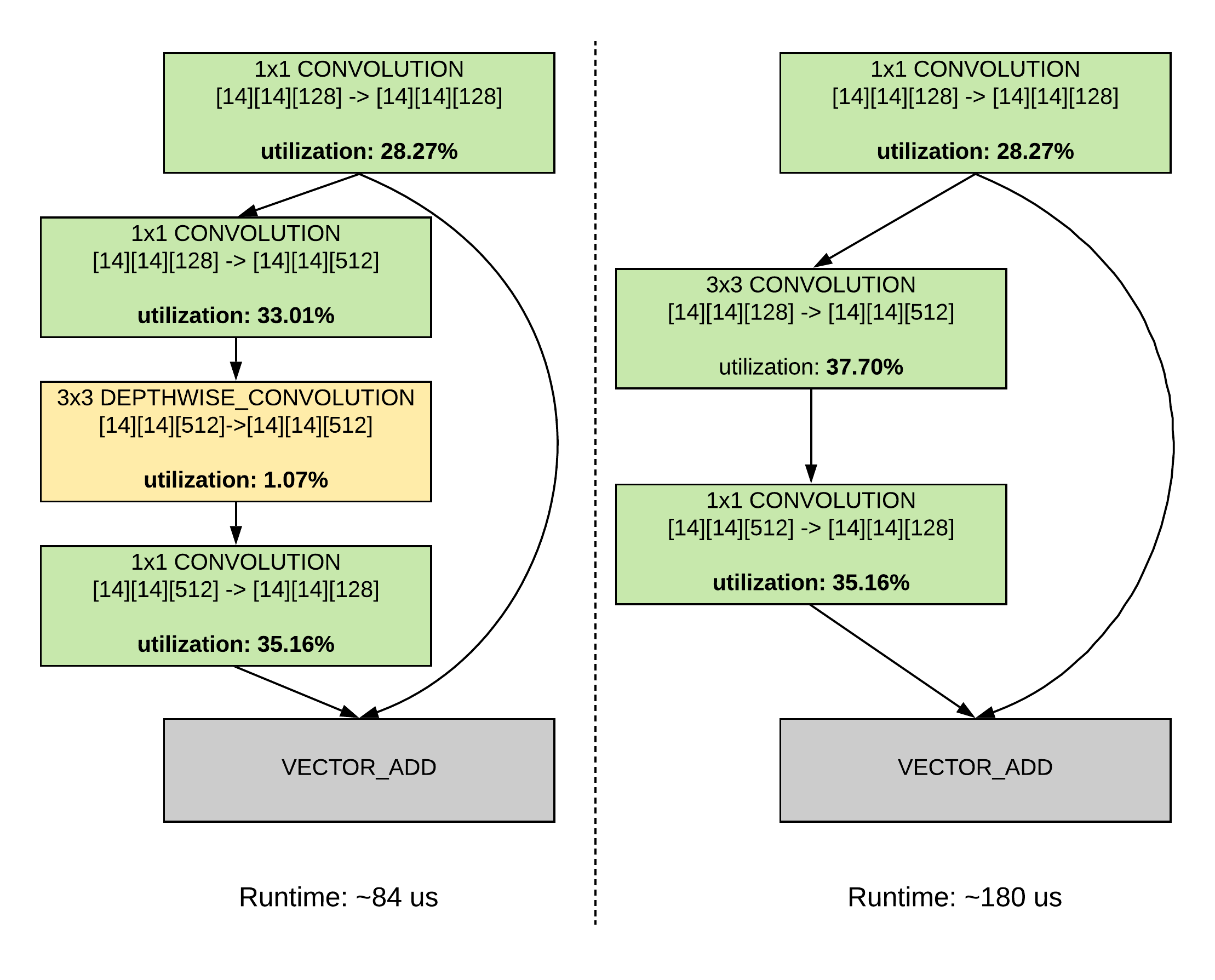}
    \vskip -0.2in
    \caption{Inverted bottleneck convolution block with and without depthwise convolution layer have significantly different runtimes depending on the input tensor dimension and block expansion factors}
        \vskip -0.1in

    \label{fig:SSE}
\end{figure}

Although our search space includes several potentially useful blocks with varying kernel and tensor sizes, it is not trivial to determine when an option becomes favorable. For example, our search space includes the inverted bottleneck convolution block with a depthwise convolution layer that is used in MobileNetV2 \cite{Sandler2018MobileNetV2IR}. In addition to this baseline block, we introduce a fused inverted bottleneck convolution block that fuses the initial expansion convolution with the depthwise convolution into a single full convolution (Figure~\ref{fig:SSE}). Originally this block expands the depth of the input tensor and performs a ``cheaper" depthwise convolution with a larger depth dimension. Although, the fused alternative performs a more ``expensive" full convolution at a larger depth dimension, it can utilize the hardware resources better and provide more trainable parameters which can be a good latency-accuracy trade-off. In Figure~\ref{fig:SSE}, on the top, we observe that the fused inverted bottleneck block has a better runtime as well as more trainable parameters compared to the baseline inverted bottleneck. However, on the bottom, fused version has more than 2x worse runtime compared to the baseline version.

\begin{figure}[t]
    \centering
    \includegraphics[width=\linewidth-2em]{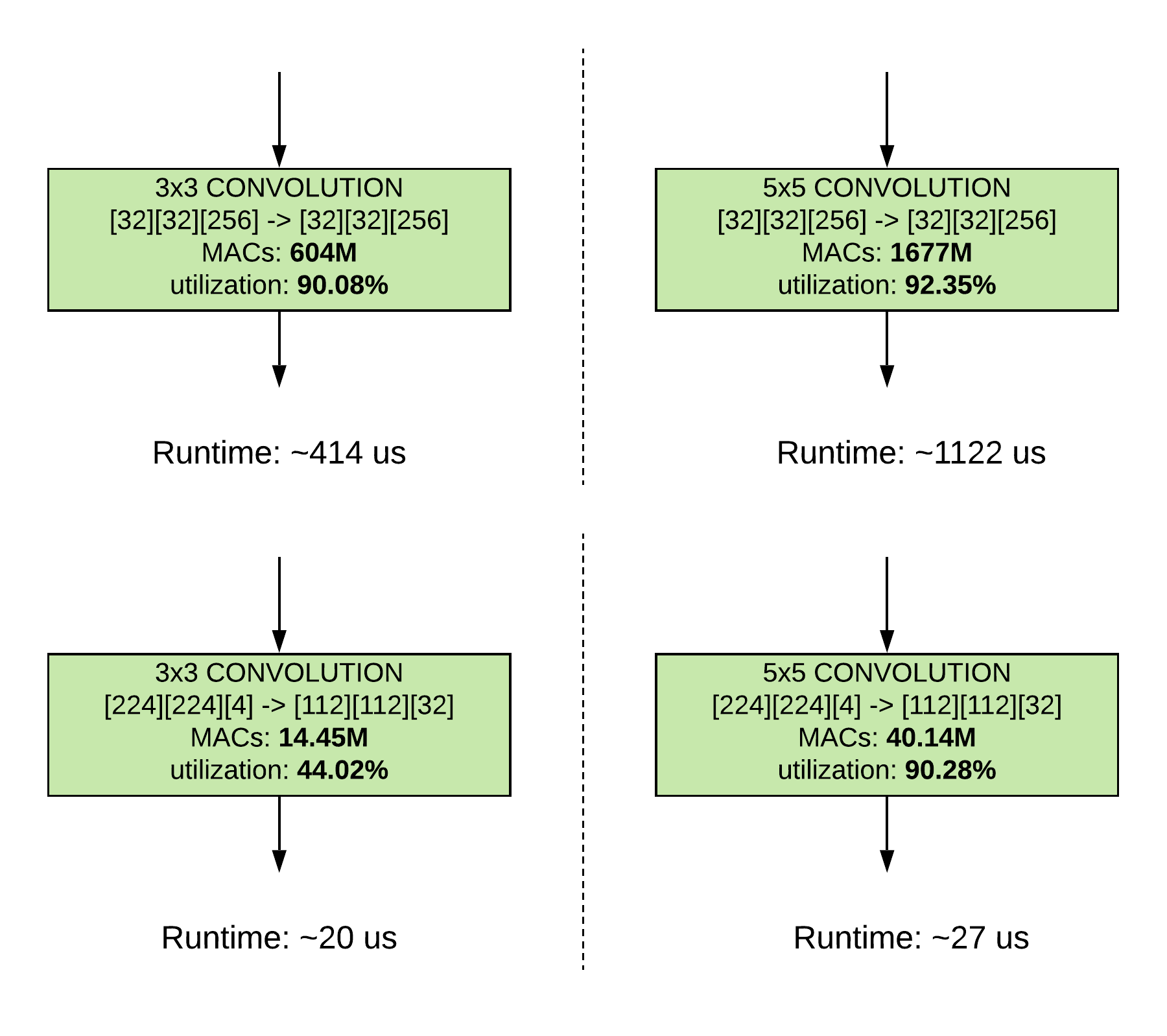}
    \vskip -0.2in
    \caption{Convolution kernel size's impact on the runtime changes significantly based on the input/output tensor shapes.}
        \vskip -0.1in

    \label{fig:kernel-size}
\end{figure}

Figure~\ref{fig:kernel-size} demonstrates another case where the same choice from the search space is not always favorable. In Figure~\ref{fig:kernel-size}, on the top, 5x5 kernel size choice leads to 2.78x increase in the number of MACs and parameters compared to 3x3 kernel size which leads to 2.71x increase in the runtime (1122us vs. 414us). However, on the bottom we observe that the same increase in the kernel size, number of MACs and parameters lead to only a 35\% increase in the runtime (27us vs 20us). For this case, it turns out that the combination of a shallow input tensor depth with a larger output tensor depth has a lower utilization where the increase in the kernel size has minor impact on the runtime due to improved utilization. This can be good trade-off to gain more trainable parameters to improve model quality at a marginal latency cost.

The cost of such choices depends on several factors including the input/output tensor shapes, where they appear in the model and how they are mapped onto the hardware. Determining the crossover point where an option becomes more beneficial in terms of the latency-accuracy trade-off is a non-trivial problem which makes manual model crafting very challenging. This makes the accelerator-aware NAS an essential approach to improve the model accuracy while efficiently utilizing the system performance.

\section{Results}

%\begin{table}[t]
%\caption{Latencies of different models on Pixel4 CPU and Edge TPU. (Imagenet Top-1 accuracy for int8-quantized models)}
%\vspace{-2mm}
%\begin{center}
%\begin{small}
%\begin{tabular}{m{7.5em}m{2em}m{2em}m{2.5em}|m{2em}m{2em}}
%\toprule
%& & & & \multicolumn{2}{c}{Latency(ms)} \\
%Model & Params & Mult-Adds & Top-1 & Pixel4 CPU  & Pixel4 Edge TPU \\
%\midrule
%MobilenetV3-Large & 5.4M & 217M  & 73.9 & 10 & N/A\\
%\vskip 0.05in
%MobilenetV3-Large minimalistic & 3.9M & 207M & 71.3 & 8.8 & 2.95 \\
%\vskip 0.05in
%\textbf{MobilenetEdgeTPU} & \textbf{4.0M} &\textbf{990M} & \textbf{75.6} & \textbf{20.6} & \textbf{3.6} \\
%\bottomrule
%\end{tabular}
%\end{small}
%\end{center}
%\vskip -0.1in
%\end{table}
%\label{tab:mobilenet}

We present two set of results targeting the Edge TPU in Coral devices and 
in Pixel 4 smartphones for a class of image classification models designed using accelerator-aware NAS.

\begin{figure}[t]
    \centering
    \includegraphics[width=\linewidth-3em]{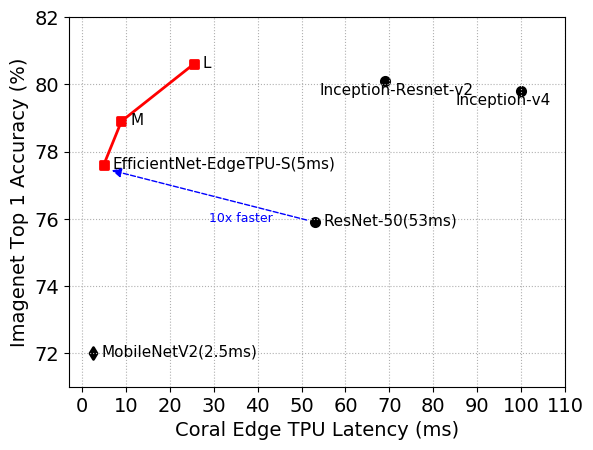}
    \vskip -0.1in
    \caption{EfficientNet-EdgeTPU-S/M/L models achieve better latency and accuracy than ResNet50, and Inception by specializing the network architecture for Edge TPU in Coral devices.}
    \label{fig:efficientnet}
    \vskip -0.1in
\end{figure}

The search space for Edge TPU is derived from the Mnasnet search space \cite{mnasnet}. This search space is customized for the Edge TPU by including the option to use the fused inverted bottleneck convolution block described previously. Also, the swish non-linearity and the squeeze-and-excite blocks \cite{Hu2018} are excluded from the search space. While these operations have been shown to help improve the model's accuracy \cite{Howard2019SearchingFM}, they tend to perform suboptimally on Edge TPU. An architecture search targeting the Edge TPU in Coral devices produced the model shown in Figure~\ref{fig:efficientnet} labelled as EfficientNet-EdgeTPU-S. The compound scaling method of EfficientNets \cite{tan2019efficientnet} is used the produce the -M and -L variants. 
The scaled versions -M and -L achieve progressively higher accuracy at the cost of higher latency.
Compared with other image classification models such as Inception-ResNet-v2 and ResNet50, EfficientNet-EdgeTPU models are not only more accurate, but also run faster on Edge TPUs.

\begin{figure}[t]
    \centering
    \includegraphics[width=\linewidth-3.2em]{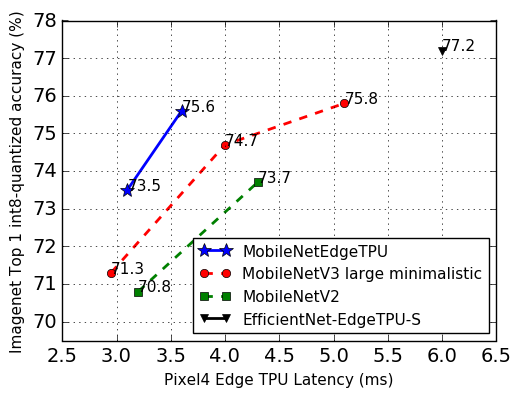}
    \vspace{-3mm}
    \includegraphics[width=\linewidth-3.2em]{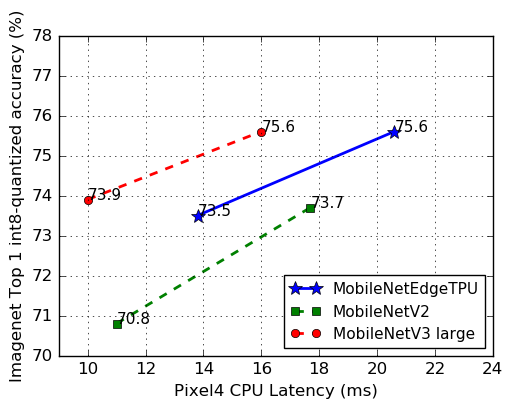}
    \vskip -0.1in
    \caption{Latencies of different variants of int8-quantized Mobilenets on Pixel 4 Edge TPU (top) and Pixel 4 CPU (bottom)}
    \label{fig:mobilenet}
    \vskip -0.1in
\end{figure}

Pixel 4 uses an Edge TPU customized to meet the requirements of key camera features in Pixel 4. 
This allows us to reuse a similar search space with only minor changes. Since the target hardware and hence the cost models are different, the accuracy-latency trade-off curves for selecting the optimal choices changes as well.
However, the accelerator-aware NAS approach substantially reduces the manual process involved in handcrafting a new optimal model. As shown in Figure~\ref{fig:mobilenet}, our NAS generated models, MobilenetEdgeTPU, improve the accuracy-latency pareto-frontier compared to existing mobile models such as MobileNetV2 and minimalistic MobileNetV3.
Compared with the EfficientNet-EdgeTPU model (optimized for the Edge TPU in Coral),
these models are targeted to run at a much lower latency on Pixel 4. 

\textbf{No One-Size-Fits-All}:
Note that the improvements demonstrated above arise due to the fact that these models have been customized to run on the Edge TPU accelerator. When running on a mobile CPU, MobileNetEdgeTPU delivers a lower performance compared to the models that have been customized for mobile CPUs (MobileNetV3) (Figure~\ref{fig:mobilenet}).
MobileNetEdgeTPU models perform a much greater number of MAC operations (990M vs. 210M), hence, it is not surprising that they run slower on mobile CPUs, which exhibit a more linear relationship between a model’s compute requirements and the runtime.

\section{Conclusions}
%Diminishing returns on transistor scaling and performance gains for general-purpose processors in the post-Moore’s Law era in tandem with a growing demand for real-time, computationally intensive machine learning fueled the growth of hardware accelerators for neural networks.
This work highlights the benefits of customization of neural network architectures for hardware accelerator architectures. Neural network design using AutoML can help substantially reduce the manual effort involved in these accelerator-specific customizations. It is evident that such customization provides a path forward for continued improvement in system performance in the post-Moore's law era.

% Acknowledgements should only appear in the accepted version.
\section*{Acknowledgements}
This work is made possible through a collaboration spanning several teams across Google. We’d like to acknowledge contributions from Gabriel Bender, Bo Chen, Andrew Howard, Eddy Hsu, John Joseph, Pieter-jan Kindermans, Quoc Le, Owen Lin, Hanxiao Liu, Yun Long, Ravi Narayanaswami, Mark Sandler, Mingxing Tan, Dong Hyuk Woo, Yunyang Xiong and support from Chidu Krishnan and Steve Molloy.

% In the unusual situation where you want a paper to appear in the
% references without citing it in the main text, use \nocite

%\bibliographystyle{sysml2019}

%%%%%%%%%%%%%%%%%%%%%%%%%%%%%%%%%%%%%%%%%%%%%%%%%%%%%%%%%%%%%%%%%%%%%%%%%%%%%%%
%%%%%%%%%%%%%%%%%%%%%%%%%%%%%%%%%%%%%%%%%%%%%%%%%%%%%%%%%%%%%%%%%%%%%%%%%%%%%%%
% SUPPLEMENTAL CONTENT AS APPENDIX AFTER REFERENCES
%%%%%%%%%%%%%%%%%%%%%%%%%%%%%%%%%%%%%%%%%%%%%%%%%%%%%%%%%%%%%%%%%%%%%%%%%%%%%%%
%%%%%%%%%%%%%%%%%%%%%%%%%%%%%%%%%%%%%%%%%%%%%%%%%%%%%%%%%%%%%%%%%%%%%%%%%%%%%%%
% \appendix
% \section{Please add supplemental material as appendix here}
% %
% Put anything that you might normally include after the references as an appendix here, {\it not in a separate supplementary file}. Upload your final camera-ready as a single pdf, including all appendices.

%%%%%%%%%%%%%%%%%%%%%%%%%%%%%%%%%%%%%%%%%%%%%%%%%%%%%%%%%%%%%%%%%%%%%%%%%%%%%%%
%%%%%%%%%%%%%%%%%%%%%%%%%%%%%%%%%%%%%%%%%%%%%%%%%%%%%%%%%%%%%%%%%%%%%%%%%%%%%%%

\end{document}